\documentclass[aps,prl,groupedaddress,showpacs,graphics]{revtex4}

\usepackage{graphics}
\usepackage{psfig}
\usepackage{epsf}

\begin{document}


\title{
Generalized Wigner Function Formulation \\
for Quantum Systems with Open Boundaries
}


\author{Remo Proietti Zaccaria}
\email[]{Proietti@Athena.PoliTo.It}
\author{Fausto Rossi}
\email[]{FRossi@Athena.PoliTo.It}
\affiliation{
Istituto Nazionale per la Fisica della Materia (INFM) and
Dipartimento di Fisica, Politecnico di Torino,
Corso Duca degli Abruzzi 24,
10129 Torino, Italy
}


\date{\today}

\begin{abstract}

A rigorous microscopic theory for the description of quantum-transport phenomena in
systems with open boundaries is proposed. We shall show that the application of the conventional Wigner-function formalism to this problem leads to unphysical results, like injection of coherent electronic states from the contacts. To overcome such basic limitation, we propose a generalization of the standard Wigner-function formulation, able to properly describe the incoherent nature of carrier injection at the device spatial boundaries as well as the interplay between phase coherence and energy relaxation/dephasing within the device active region. The proposed theoretical scheme constitutes the first rigorous derivation of the phenomenological injection model commonly employed in  the simulation of open quantum devices.

\end{abstract}

\pacs{
72.10.Bg, 85.30.-z, 73.40.-c
}

\maketitle



Present-day technology pushes device dimensions toward limits
where the traditional semiclassical transport theory~\cite{MC} can
no longer be applied, and more rigorous quantum-transport
approaches are required~\cite{QT}. However, in spite of the
quantum-mechanical nature of carrier dynamics in the core region
of typical nanostructured devices ---like semiconductor
superlattices and double-barrier structures--- the overall
behaviour of such quantum systems is often the result of a complex
in\-ter\-play between phase coherence and energy
relaxation/dephasing~\cite{QCL}, the latter being primarily due to
the presence of spatial boundaries~\cite{WF}. It follows that a
proper treatment of such novel nanoscale devices requires a
theoretical modelling able to properly account for both coherent
and incoherent ---i.e., phase-breaking--- processes on the same
footing. To this end, a generalization to open systems ---i.e.,
systems with open boundaries--- of the well known Semiconductor
Bloch Equations (SBE) \cite{SBE} has been recently
proposed~\cite{PRL}. However, the theoretical analysis presented
in \cite{PRL} is primarily related to the interplay between phase
coherence and energy relaxation within the device active region,
and
---apart from its abstract formulation--- no detailed
investigation of the carrier-injection process (from the
electrical contacts into the device active region) has been
performed so far.

Aim of the present Letter is to provide a rigorous quantum-mechanical description of the coupling dynamics between the device active region and external charge reservoirs, able to account for the semi-phenomenological injection models commonly employed in
state-of-the-art simulations of realistic one- and two-dimensional open quantum
devices~\cite{IWCE}.
Among such simulation strategies it is worth mentioning the approach recently proposed by Fischetti and co-workers~\cite{Max,note-Max}:
By denoting with $f_\alpha$ the carrier distribution over the electronic states $\alpha$ of the device, the transport equation proposed in \cite{Max} is of the form:
\begin{equation}\label{e:phen-mod}
\frac{d}{dt} f_{\alpha} = \sum_{\alpha'}
\left( W_{\alpha\alpha'} f_{\alpha'} - W_{\alpha'\alpha} f_\alpha \right) + \frac{f^b_\alpha - f_\alpha}{\tau_\alpha}\ .
\end{equation}
Here, $f^b_\alpha$ denotes the equilibrium carrier distribution in the contacts while $\tau_\alpha$ can be regarded as the device transit time for an electron in state $\alpha$.
As anticipated, in spite of a rigorous treatment of the scattering dynamics,
the last (relaxation-time-like) term
describes carrier injection/loss on a partially phenomenological level and does not depend on the real position of the device spatial boundaries.

In order to provide a fully microscopic real-space description
of the carrier-injection process, we shall start rivisiting the theoretical approach proposed in \cite{PRL}.
By applying to the conventional SBE for a closed system~\cite{PRL,note-rho,note-L}
\begin{equation}\label{e:SBE}
\frac{d}{dt} \rho_{\alpha_1\alpha_2} = \sum_{\alpha_1'\alpha_2'}
L_{\alpha_1\alpha_2,\alpha_1'\alpha_2',} \rho_{\alpha_1'\alpha_2'}
\end{equation}
the usual Weyl-Wigner transform
\begin{equation}\label{e:u}
u_{\alpha_1\alpha_2}({\bf r,k}) = \int d{\bf r}'
\phi_{\alpha_1}\left({\bf r}+{{\bf r}'\over 2}\right) {e^{-i{\bf k
\cdot r}'} \over (2\pi)^{3 \over 2}} \phi^*_{\alpha_2}\left({\bf
r}-{{\bf r}'\over 2}\right)
\end{equation}
one gets:
\begin{equation}\label{e:WFE1}
\frac{d}{dt} f({\bf r,k}) = \int L({\bf r,k;r',k'}) f({\bf r',k'})
d{\bf r'}\, d{\bf k'} \ ,
\end{equation}
where
\begin{equation}\label{e:WF}
f({\bf r,k}) = \sum_{\alpha_1\alpha_2} u_{\alpha_1\alpha_2}({\bf r,k}) \rho_{\alpha_1\alpha_2}
\end{equation}
is the well-known Wigner function~\cite{WF,Buot} while the quantity $L({\bf
r,k;r',k'})$ is the Liouville operator in the new phase-space
representation ${\bf r,k}$.
\begin{figure}
\unitlength1mm \ \par\noindent
\begin{picture}(160,140)
\put(0,-56.){\psfig{figure=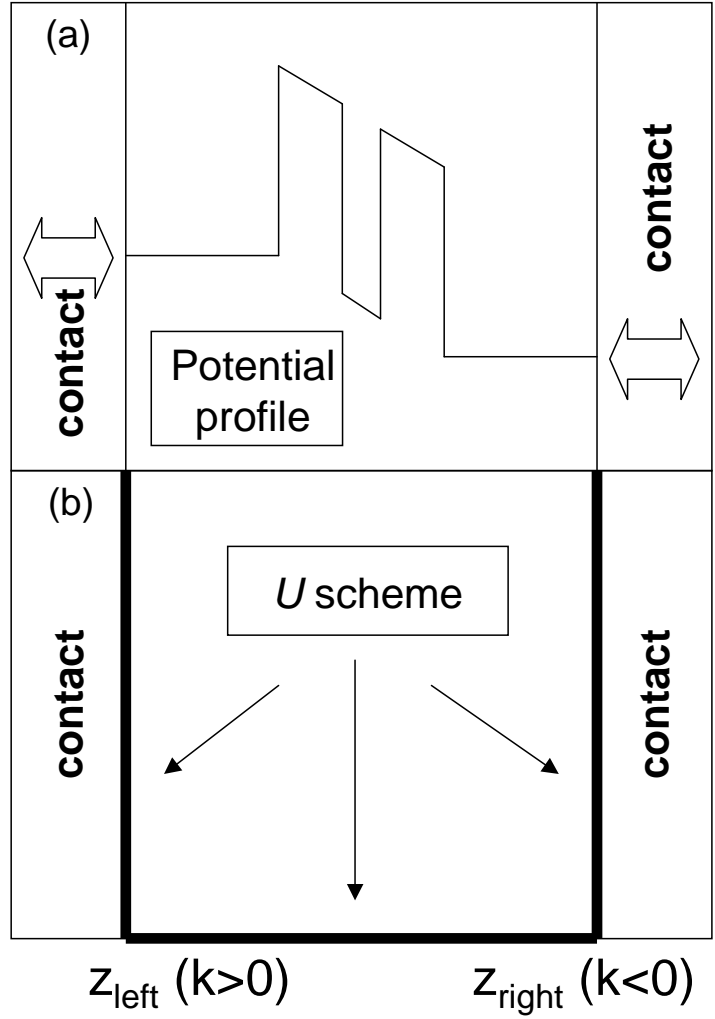,width=160mm}}
\end{picture}
\caption{ Schematic representation of the    device active region
sandwiched between its electrical contacts (a) and of the
corresponding ${\bf U}$ boundary-condition scheme for a
one-dimensional system. The latter implies, in particular, the
knowledge of the incoming Wigner function $f(z_b,k)$, i.e.,
$f(z_{\mbox{left}},k > 0)$ and $f(z_{\mbox{right}},k < 0)$.
\label{fig1}}
\end{figure}

For a closed system, the Wigner function $f$ is defined for any
value of the real-space coordinate ${\bf r}$ and its time
evolution is fully dictated by its initial condition. In contrast,
for a system with open boundaries $f$ is defined only within a
given region $\Omega$ of interest and its time evolution is determined by
the initial condition at $t_\circ$ inside such region as well as
by the value of its incoming part on the boundary ${\bf r}_b$ of the domain at any
time $t' > t_\circ$. Such boundary-condition scheme~\cite{Modena}
---also referred to as ``{\bf U} scheme''--- is depicted in
Fig.~\ref{fig1}.
More specifically, in order to properly impose the desired spatial boundary conditions, let us rewrite Eq.~(\ref{e:WFE1}) as
\begin{equation}\label{e:WFE2}
\frac{d}{dt} f({\bf r,k}) = \int \tilde{L}({\bf r,k;r',k'}) f({\bf
r',k'}) d{\bf r'} d{\bf k'} + S({\bf r,k})
\end{equation}
by adding and subtracting a source term
$S({\bf r,k}) = v({\bf k}) f^b({\bf k}) \delta({\bf r-r}_b)$ \cite{note-S}.
The latter is compensated by a renormalization $\Delta L = \tilde{L} - L$ of
the Liouville operator $L$ in (\ref{e:WFE1}) \cite{note-sbc}:
\begin{equation}\label{e:DeltaL}
\Delta L({\bf r,\!k;\!r',\!k'}) =
- v({\bf k}) \delta({\bf r-r}_b) \delta({\bf r-r'})
\delta({\bf k-k'})\ .
\end{equation}
By applying the inverse of the Weyl-Wigner transform in (\ref{e:u}) to Eq.~(\ref{e:WFE2}) we get:
\begin{equation}\label{e:SBE-open}
\frac{d}{dt} \rho_{\alpha_1\alpha_2} =
\sum_{\alpha_1'\alpha_2'}
\tilde{L}_{\alpha_1\alpha_2,\alpha_1'\alpha_2'} \rho_{\alpha_1'\alpha_2'}
+ S_{\alpha_1\alpha_2}\ .
\end{equation}
Equation (\ref{e:SBE-open}) is the desired generalization to open systems of the SBE in (\ref{e:SBE}) \cite{note-oc}.

\begin{figure}
\unitlength1mm \ \par\noindent
\begin{picture}(160,140)
\put(12,0.){\psfig{figure=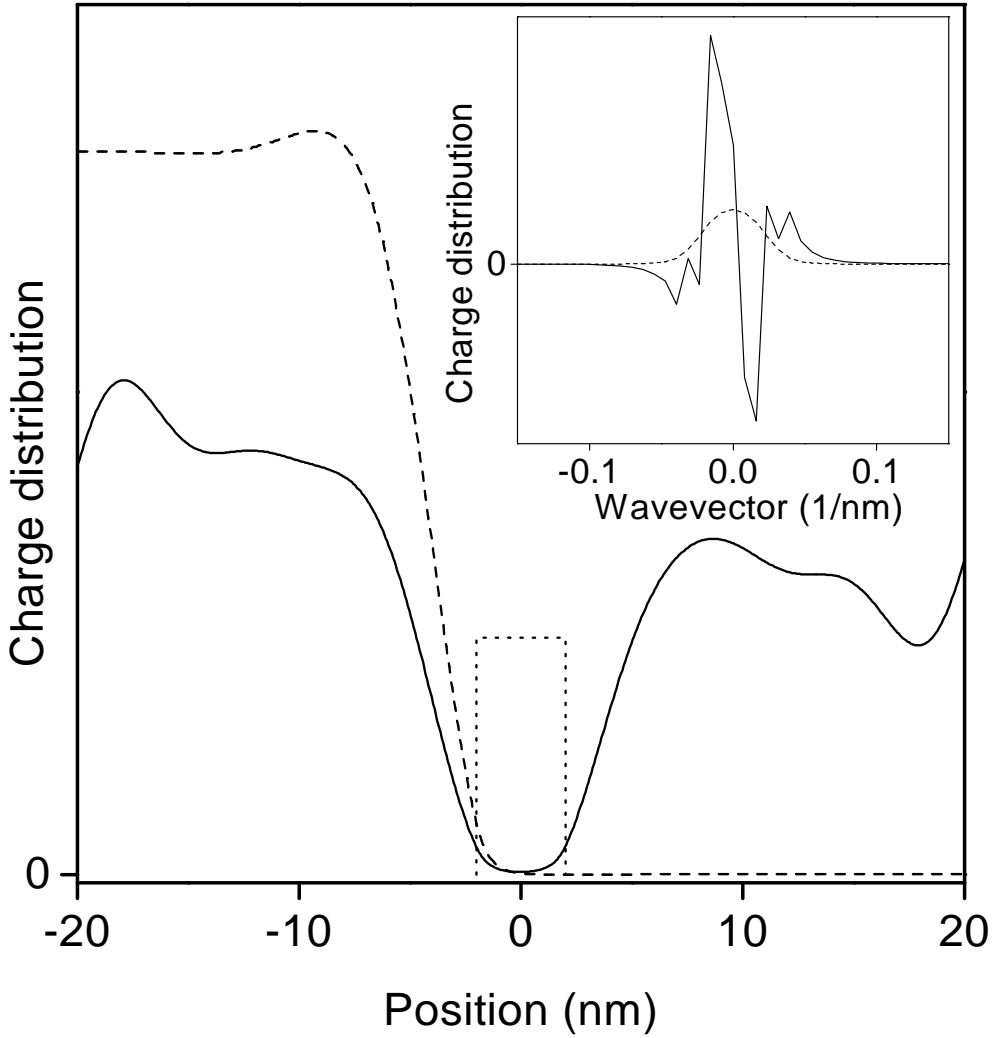,width=130mm}}
\end{picture}
\caption{
Comparison between the real-space charge distribution obtained from the phenomenological injection model in (\ref{e:phen-mod})
[$n({\bf r}) = \sum_\alpha f_\alpha \vert\phi_\alpha({\bf r})\vert^2$ --- dashed curve]
and the microscopic model in (\ref{e:ss})
[$n({\bf r}) = \sum_{\alpha_1\alpha_2} \rho_{\alpha_1\alpha_2} \phi_{\alpha_1}({\bf r}) \phi^*_{\alpha_2}({\bf r})$ --- solid curve]
for a GaAs-based single-barrier structure (height $V_\circ = 0.5$\,eV and width $a = 4$\,nm) equidistant from the electrical contacts.
In this room-temperature simulation, due to a misalignment $\Delta\mu = 0.2$\,eV of the left and right chemical potential, carriers are primarily injected from left.
The corresponding charge distribution in momentum space is also reported in the inset (see text). \label{fig2}}
\end{figure}

In order to validate the theoretical approach presented so far, we shall focus on a very simple semiconductor nanostructure: a single-barrier equidistant from the device contacts (see Fig.~\ref{fig2}). As basis states $\alpha$ we adopt the scattering states of the device potential profile; moreover, to better identify the role played by carrier injection, we shall neglect all other sources of energy relaxation/dephasing in the device active region, like carrier-phonon and carrier-carrier scattering.
Under these assumptions, Eq.~(\ref{e:SBE-open}) in steady-state conditions reduces to:
\begin{equation}\label{e:ss}
{i \over \hbar} \left(\epsilon_{\alpha_1}-\epsilon_{\alpha_2}\right) \rho_{\alpha_1\alpha_2}
-\sum_{\alpha_1'\alpha_2'}
\Delta L_{\alpha_1\alpha_2,\alpha_1'\alpha_2'} \rho_{\alpha_1'\alpha_2'}
= S_{\alpha_1\alpha_2}\ ,
\end{equation}
where $\epsilon_\alpha$ denotes the energy of the scattering state $\alpha$ and $\Delta L_{\alpha_1\alpha_2,\alpha'_1\alpha'_2}$ is the renormalization term in (\ref{e:DeltaL}) written in the density-matrix representation $\alpha_1\alpha_2$.

Figure \ref{fig2} shows results for the single-barrier potential profile when carriers are primarily injected from left. Here, the simulated real-space charge distribution obtained from the phenomenological injection model in Eq.~(\ref{e:phen-mod}) (dashed curve) is compared to that of the microscopic model in (\ref{e:ss}) (solid curves).
As we can see, the two models give completely different results. The phenomenological model gives basically what we expect: since we have significant carrier injection from left only and since the potential barrier is relatively high, the carrier distribution is mainly located on the left side. In contrast, the microscopic model gives an almost symmetric charge distribution.
In order to understand the origin of this unphysical result, let us focus on the nature of the source term in (\ref{e:SBE-open}). Contrary to the phenomenological injection/loss term in (\ref{e:phen-mod}), the latter is intrinsically non-diagonal, i.e., the injection of a carrier with well-defined wavevector ${\bf k}$ is described by a non-diagonal source contribution $S_{\alpha_1\alpha_2}$. In other words, {\it we inject into the device active region a coherent superposition of states $\alpha_1$ and $\alpha_2$},
in clear contrast with the idea of injection from a thermal ---i.e., diagonal--- charge reservoir.
More specifically, in this case the generic scattering state $\alpha$ on the left comes out to be an almost equally weighted superposition of $+k$ and $-k$:
$\phi_k(z) = a_k e^{ikz} +b_k e^{-ikz}$. This, in turn, tells us that the generic plane-wave state $k$ injected from the left contact is also an almost equally weighted superposition of the left and right scattering states. This is why the charge distribution (solid curve in Fig.~\ref{fig2}) is almost symmetric: any electron injected from left couples to left as well as to right scattering states.
The anomaly of the microscopic model is even more pronounced if we look at the carrier distribution in momentum space (see inset in Fig.~\ref{fig2}). While for the phenomenological model (dashed curve) we get a positive-definite distribution showing, as expected, the two symmetric wavevector components of the scattering state, the microscopic result is
not positive definite; this tells us that our boundary-condition scheme does not ensure that the solution of Eq.~(\ref{e:WFE2}) is a ``good'' Wigner function.

The scenario previously discussed is highly non-phy\-si\-cal; it
can be mainly ascribed to the boundary-condition scheme employed
so far, which implies injection of plane-wave electrons [see
source term in Eq.~(\ref{e:WFE2})], regardless of the shape of the
device potential profile. This is an intrinsic limitation of the
conventional Wigner-function representation ${\bf r,k}$. It is
then clear that, in order to overcome this limitation, what we
need is {\it a boundary-condition scheme realizing diagonal
injection over the scattering states $\alpha$ of the device
potential profile}.

To this end, in this Letter we propose a generalization of the Wigner-function formulation considered so far. The key idea is to extend the Weyl-Wigner transform in (\ref{e:u}) from the ${\bf k}$ to a generic basis set $\{\vert \beta \rangle\}$
according to~\cite{note-u}:
\begin{widetext}
\begin{equation}\label{e:u-new}
u^{\alpha_1\alpha_2}_{\beta_1\beta_2}({\bf r}) = \Omega \int d{\bf r}'
\phi^{ }_{\alpha_1}\left({\bf r}+{{\bf r}'\over 2}\right)
\chi^*_{\beta_1}\left({\bf r}+{{\bf r}'\over 2}\right)
\chi^{ }_{\beta_2}\left({\bf r}-{{\bf r}'\over 2}\right)
\phi^*_{\alpha_2}\left({\bf r}-{{\bf r}'\over 2}\right) \ .
\end{equation}
\end{widetext}
In analogy to (\ref{e:WF}), our generalized Wigner function is given by~\cite{note-WF-new}:
\begin{equation}\label{e:WF-new}
f_{\beta_1\beta_2}({\bf r}) = \sum_{\alpha_1\alpha_2} u_{\beta_1\beta_2}^{\alpha_1\alpha_2}({\bf r}) \rho_{\alpha_1\alpha_2}\ .
\end{equation}
By adopting as basis states $\vert \beta \rangle$ the scattering states of the device potential profile and assuming a diagonal source term of the form
\begin{equation}\label{e:S-new}
S_{\beta_1\beta_2}({\bf r}) = v_{\beta_1} f^b_{\beta_1}
\delta_{\beta_1\beta_2} \delta({\bf r-r}_b)\ ,
\end{equation}
 its equation of motion will be
\begin{widetext}
\begin{equation}\label{e:WFE2-new}
\frac{d}{dt} f_{\beta_1\beta_2}({\bf r}) = \sum_{\beta'_1\beta'_2} \int d{\bf r'} \tilde{L}_{\beta_1\beta_2,\beta'_1\beta'_2}({\bf r,r'}) f_{\beta'_1\beta'_2}({\bf
r'}) + S_{\beta_1\beta_2}({\bf r})
\end{equation}
\end{widetext}
with a renormalization $\Delta L_{\beta_1\beta_2,\beta'_1\beta'_2}({\bf r,r'})$ given by
\begin{equation}\label{e:DeltaL-new}
-
v_{\beta_1}  \delta_{\beta_1,\beta_2}
\delta_{\beta_1\beta_2,\beta'_1\beta'_2} \delta({\bf r-r}_b)
\delta({\bf r-r'})\ .
\end{equation}
We stress that now the source term in (\ref{e:S-new}) describes diagonal injection over the scattering states (with velocity $v_\beta$), as requested.
Indeed, if we now integrate Eq.~(\ref{e:WFE2-new}) over the real-space coordinate ${\bf r}$, we get again the density-matrix equation in (\ref{e:SBE-open}), but with a diagonal source term
$S_{\alpha_1\alpha_2} = v_{\alpha_1} f^b_{\alpha_1}
\delta_{\alpha_1\alpha_2}$
and a much simpler ---i.e., partially diagonal---
renormalization term
$\Delta L_{\alpha_1\alpha_2,\alpha'_1\alpha'_2} = -v_{\alpha_1} \delta_{\alpha_1\alpha_2} u^{\alpha'_1\alpha'_2}_{\alpha_1\alpha_2}({\bf r}_b)$.
In the scattering-free case, the stationary solution is again
described by Eq.~(\ref{e:ss}). However, due to the diagonal nature
of the new source term as well as of the partially diagonal structure
of
$\Delta L$, Eq.~(\ref{e:ss}) has now a
diagonal solution: $\rho_{\alpha_1\alpha_2} = f_{\alpha_1}
\delta_{\alpha_1\alpha_2}$.
More specifically, the diagonal density-matrix elements $f_\alpha$ obey
the following steady-state equation:
\begin{equation}\label{e:ss-new}
{\cal T}_{\alpha\alpha'} f_{\alpha'} = f^b_\alpha
\end{equation}
with ${\cal T}_{\alpha\alpha'} = u^{\alpha\alpha}_{\alpha'\alpha'}({\bf r}_b)$. Equation
(\ref{e:ss-new}) is semiclassical in nature, i.e., it involves
diagonal density-matrix terms only~\cite{note-limit}.
\begin{figure}
\unitlength1mm \ \par\noindent
\begin{picture}(160,140)
\put(12,0.){\psfig{figure=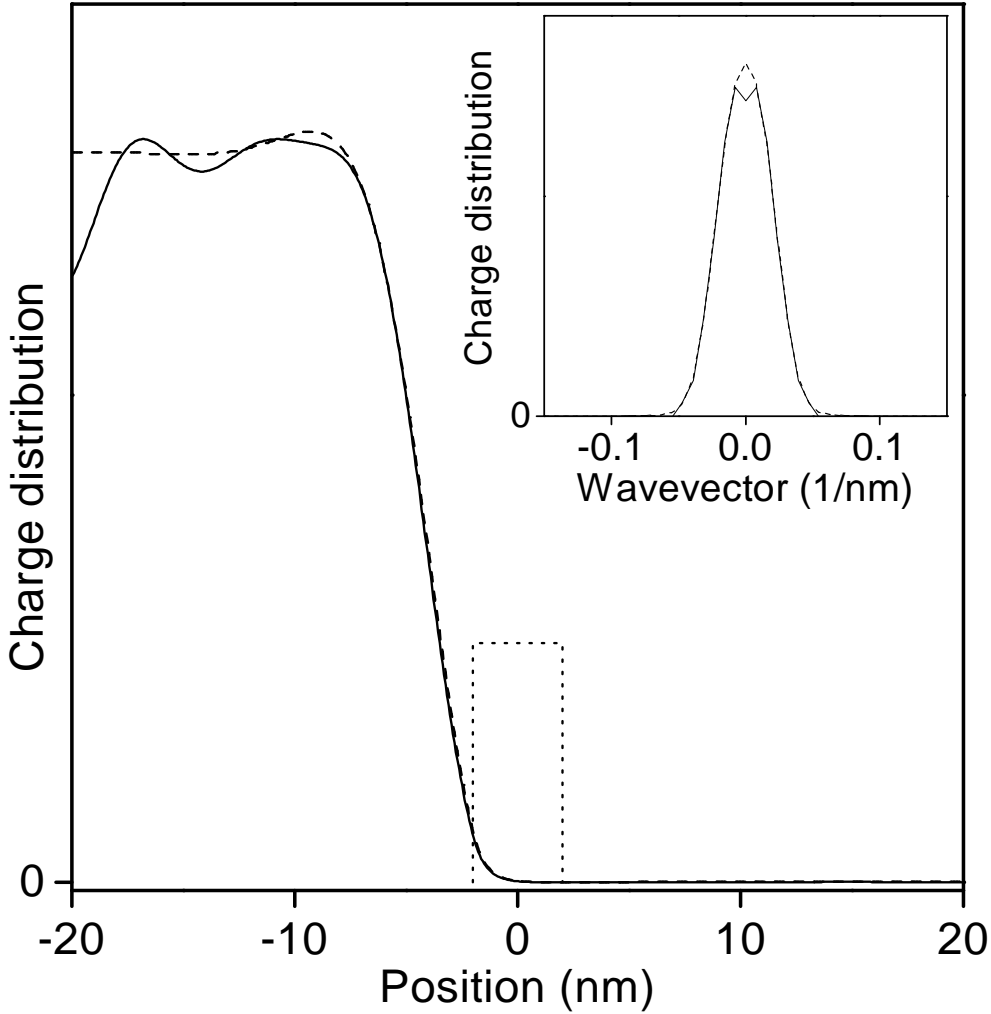,width=130mm}}
\end{picture}
\caption{
Same as in Fig.~\ref{fig2} but for the new microscopic model in Eq.~(\ref{e:ss-new}) (see text).
\label{fig3}}
\end{figure}

Figure \ref{fig3} shows again results for the single-barrier potential profile previously considered. Here, the simulation based on the phenomenological injection model in Eq~(\ref{e:phen-mod}) (dashed curves) is compared to that of the new microscopic model in (\ref{e:ss-new}) (solid curves).
As we can see, the highly non physical behaviours of Fig.~\ref{fig2} (solid curves) have been completely removed. Indeed, the momentum distribution in the inset
is always positive-definite and the two models exhibit a very similar behaviour.
We find relatively small deviations close to the device spatial boundaries,
which can be ascribed to the interlevel injection coupling ${\cal T}_{\alpha\alpha'}$ [see Eq.~(\ref{e:ss-new})], not present in the phenomenological injection model.

In conclusion, we have proposed a rigorous description of quantum-transport phenomena in
systems with open boundaries. Our analysis has shown that the conventional Wigner-function formalism leads to unphysical results, like injection of coherent superpositions of states from the device spatial boundaries.
This basic limitation has been removed by introducing a generalization of the standard Wigner-function formulation, able to properly describe the incoherent nature of carrier injection. The proposed theoretical scheme constitutes a rigorous derivation of the phenomenological injection models commonly employed in  the simulation of open quantum devices.


\begin{acknowledgments}

We are grateful to Aldo Di Carlo, Massimo Fischetti, Rita Iotti, Carlo Jacoboni, Paolo Lugli, and Paolo Zanardi for a number of stimulating and fruitful discussions.

\end{acknowledgments}


\begin{thebibliography}{xx}
\bibitem{note-Max}
This is grounded on a Master Equation description derived from a rigorous density-matrix formulation of the problem, and can be regarded as the diagonal (i.e., semiclassical) limit of the theoretical scheme proposed in \cite{PRL}.
\bibitem{note-rho}
Here, $\rho_{\alpha_1\alpha_2}$ is the single-particle density matrix written in our single-particle basis $\{\phi_\alpha\}$. Its diagonal elements correspond to the
usual distribution function $f_\alpha$
of the semiclassical theory \protect\cite{MC}\protect
while the off-diagonal terms ($\alpha\ne\beta$) describe
the degree of quantum-mechanical phase coherence
between states
$\alpha$ and $\beta$.
\bibitem{note-oc}
The open character of the system results in a
non-Hermitian correction $\Delta L$ to the Liouville operator $L$,
whose effect is equivalent to a purely dissipative process within the
simulated region, as originally pointed out in \cite{WF}.
\bibitem{note-limit}
We stress that, by replacing the ${\cal T}$ with the identity operator
(${\cal T}_{\alpha\alpha'} = \delta_{\alpha\alpha'}$),
the usual phenomenological model in (\ref{e:phen-mod}) is recovered; indeed for the scattering-free case ($W_{\alpha\alpha'} = 0$) in steady-state conditions we get
$f_\alpha = f^b_\alpha$.
\end{thebibliography}
\end{document}